\documentclass[fleqn,usenatbib]{mnras}

\usepackage{newtxtext,newtxmath}
\usepackage{xspace}

\usepackage[T1]{fontenc}
\usepackage{ae,aecompl}
\usepackage{url}

\usepackage{graphicx}	
\usepackage{amsmath}	
\usepackage{amssymb}	
\usepackage{float}

\usepackage[utf8]{inputenc}

\title[The redshift of 4FGL~J0955.1+3551]{The redshift and the host galaxy of the neutrino candidate 4FGL~J0955.1+3551 (3HSP~J095507.9+355101)}

\author[S. Paiano et al.]{Simona Paiano$^{1,2}$\thanks{E-mail:
simona.paiano@inaf.it}, Renato Falomo$^{3}$, Paolo Padovani$^{4,5}$, Paolo Giommi$^{6,7,8}$, 
\newauthor Adriana Gargiulo$^{2}$, Michela Uslenghi$^{2}$, Andrea Rossi$^{9}$, Aldo Treves$^{10,11}$\\    
$^{1}$INAF - Osservatorio Astronomico di Roma, via Frascati 33, I-00040, Monteporzio Catone, Italy \\
$^{2}$INAF - IASF Milano, via Corti 12, I-20133, Milano, Italy \\
$^{3}$INAF - Osservatorio Astronomico di Padova, vicolo dell'Osservatorio 5, I-35122, Padova, Italy\\
$^{4}$European Southern Observatory, Karl-Schwarzschild-Str. 2, D-85748 Garching bei M\"unchen, Germany\\
$^{5}$Associated to INAF - Osservatorio Astronomico di Roma, via Frascati 33, I-00040 Monteporzio Catone, Italy\\
$^{6}$Institute for Advanced Study, Technische Universit{\"a}t M{\"u}nchen, Lichtenbergstrasse 2a, D-85748 Garching bei M\"unchen, Germany\\
$^{7}$Associated to Agenzia Spaziale Italiana, ASI, via del Politecnico s.n.c., I-00133 Roma Italy \\
$^{8}$ICRANet, Piazzale della Repubblica 10, I-65122, Pescara,
Italy\\
$^{9}$INAF - Osservatorio di Astrofisica e Scienza dello Spazio, via Piero Gobetti 93/3, 40129 Bologna, Italy\\
$^{10}$Universita' dell'Insubria, via Valeggio, 22100, Como, Italy\\
$^{11}$INAF - Osservatorio Astronomico di Brera, via Bianchi 46, I-23807, Merate (Lecco), Italy\\
}

\date{Received:~\today; Accepted:~ }

\begin{document}
\label{firstpage}
\pagerange{\pageref{firstpage}--\pageref{lastpage}}
\maketitle

\begin{abstract}
The BL Lac object 4FGL~J0955.1+3551 has been suggested as a possible source of ultra energetic neutrinos detected by the Icecube observatory. 
The target was observed in January 2020 at the Large Binocular Telescope. 
Our spectroscopy (4100-8500~$\textrm{\AA}$) yields a firm redshift z~=~0.557 as deduced by the absorption lines of the host galaxy. 
The upper limit of the minimum equivalent width on emission lines is $\sim $0.3~$\textrm{\AA}$.
From the source image we are able to resolve the host galaxy for which we measure an absolute magnitude M(R)~=~-22.9 and R$_{e}$~=~8~kpc, that is values which are typical of the host galaxies of BL Lacs.
\end{abstract}


\section{Introduction}

BL Lac objects (BLL) are the dominant population of the extragalactic $\gamma$-ray sky, both in the 0.1~--~50~GeV band, where the \textit{Fermi} mission is sensitive, and at 100~--~1000~GeV, where the data are provided by atmospheric Cherenkov telescopes. 
At higher energies and redshift z$>$0.1 direct photon detection is impeded by electron/positron pair production with photons of the extragalactic background light (EBL).
The basic picture of BLL invokes the presence of relativistic jets in the observer direction. The modeling of the sources requires the knowledge of their distance, which in the case of BLL, is hard to establish, because by definition of the class the spectral lines are absent or very weak (equivalent width, EW $<$~5~$\textrm{\AA}$). 
In the past several years some of us have been studying optical spectra of BLL with resolution and signal-to-noise ratio (S/N) optimized to detect such weak features \citep{landoni2014,paiano2017tev,paiano20173fgl,paiano2017ufo1,paiano2019ufo2}.

A major progress on BLL research has been the recent association of high energy (PeV) neutrino events of astrophysical origin  detected by the IceCube Collaboration \citep[e.g.][and references therein]{icecube2018txs,icfermi, magic2018txs,Giommi_2020b} with BLL,
which are a blazar sub-class. 
These neutrinos permit to shed light on the most energetic processes, which, as mentioned above, are otherwise screened off by the EBL.
We have therefore started a new program to determine the redshifts of BLL associated with high energy neutrinos. 
Our first target has been TXS~0506+056, which is the prototype of neutrino emitting BLL \citep{icecube2018txs,icfermi,Padovani_2018}. 
The result was the detection of very weak [O~II], [O~III] and [N~II] emission lines (EW $\sim$0.1~$\textrm{\AA}$), which established the redshift at z~=~0.3365 \citep{paiano2018txs}. 
Redshifts of other neutrino candidate sources have also been reported by some of us \citep{paiano2018ATel0627,paiano2019ATel4c41,paiano2019ATel2255}.

Here we consider the case of the $\gamma$-ray source 4FGL~J0955.1+3551 (3HSP~J095507.9+355101, named J0955 in the following). 
This is a blazar (g$\sim$20.4) that is considered a possible neutrino counterpart as it was found inside the error box of the IceCube neutrino event 200107A detected on 2020~January~07~\citep[GCN-\#26655,][]{icecube2020GCN0955}. 
Triggered by this event, the \textit{Swift} observatory was pointed at the source and revealed it at its highest ever X-ray flux \citep[Atel\#13394,][]{giommi2020ATel0955}. 
The analysis of multifrequency data shows that the synchrotron peak is at $\nu_{\rm peak}^S$ $\sim 5\times 10^{17} - 2 \times~10^{18}$ Hz making J0955 a member of the rare class of extreme blazars \citep[see][for a very recent review]{Biteau_2020}.
Details on the source, its nature and multiwavelength spectral energy distribution, and a theoretical interpretation are given in the accompanying letter by \cite{Giommi_2020c}.

The only literature optical spectrum of the object is provided by the SDSS survey.
Based on this very poor signal-to-noise (S/N$\sim$5) spectrum \citet{plotkin2008} tentatively classified the source as BL Lac object and suggested a possible redshift z~=~0.557 due to moderate host galaxy contamination. 
However no specific line identifications are provided. 

In this letter we report an optical spectroscopy and imaging study of the target which aims to determine the redshift and the nature and photometric properties of the host galaxy.
We assume in this work the cosmological parameters H$_0=$ 70 km s$^{-1}$ Mpc$^{-1}$, $\Omega_{\Lambda}$=0.7, and $\Omega_{m}$=0.3.

\section{Observations and Data Analysis}

On January 29, 2020 we observed the optical counterpart of J0955 at the Large Binocular Telescope (LBT; two twin 8.4m telescopes at Mt Graham in Arizona), using the MultiObject Double Spectrographs MODS-1 and MODS-2 \citep{pogge2010} in dual grating mode (grisms G400L and G670L), using the slit width of 1.2$^{\prime\prime}$, a dichroic at 5600~$\textrm{\AA}$, covering the spectral ranges 4100-5500~$\textrm{\AA}$ and 5750-8500~$\textrm{\AA}$, and yielding spectral resolution R~=~$\lambda/\Delta\lambda\sim$1000.
For each setting, we obtained 14 independent exposures of 900 seconds for a total exposure time of 3.5 hours.

Spectroscopy data reduction was carried out at the Italian LBT Spectroscopic Reduction Center adopting the standard procedure for long-slit spectroscopy with bias subtraction, flat-fielding, and bad-pixel correction. Each spectrum was cleaned of cosmic-ray contamination using the L.A. Cosmic algorithm \citep{vandokkum2001lacos}.
Wavelength calibration was obtained using the spectra of
Hg, Ar, Xe, and Kr lamps, providing an accuracy of $\sim$0.1~$\textrm{\AA}$ over the whole spectral range.
Relative flux calibration was derived from the observations of spectro-photometric standard stars and we anchored the
absolute flux to the magnitude of the source from acquisition images.
Finally, the spectrum was dereddened by applying the extinction law of \cite{cardelli1989dered}, assuming E(B–V)~=~0.01 as from the NASA/IPAC Infrared Science Archive 6.7

In addition to the spectroscopy data, we obtained a short exposure image (60~s) in r band.
In order to derive the properties of the host galaxy  we performed a two-dimensional fit of the LBT image using the Astronomical Image Decomposition and Analysis (AIDA) package \citep{uslenghi2008aida}.  
AIDA is an IDL-based software specifically designed to provide a simultaneous decomposition into the nucleus and the surrounding nebulosity. 
Details can be found in our previous studies of QSO host galaxies
\citep[see e.g. ][]{falomo2008,falomo2014,kotilainen2007,kotilainen2009,decarli2012}.
We then used AIDA to characterize the host galaxy properties from a 2D fit of the source adopting a model including the nucleus (modeled by the point spread function [PSF]) and the host galaxy assuming a Sersic profile convolved with the proper PSF. 
The PSF has been derived by fitting four stars with a 2D analytical model composed by a superposition of gaussians and exponential functions.
The core of the PSF has a FWHM of 0.65$^{\prime\prime}$.
\section{Results}

\subsection{Spectroscopy}

\begin{figure*}
\includegraphics[width=0.52\textwidth, angle=-90]{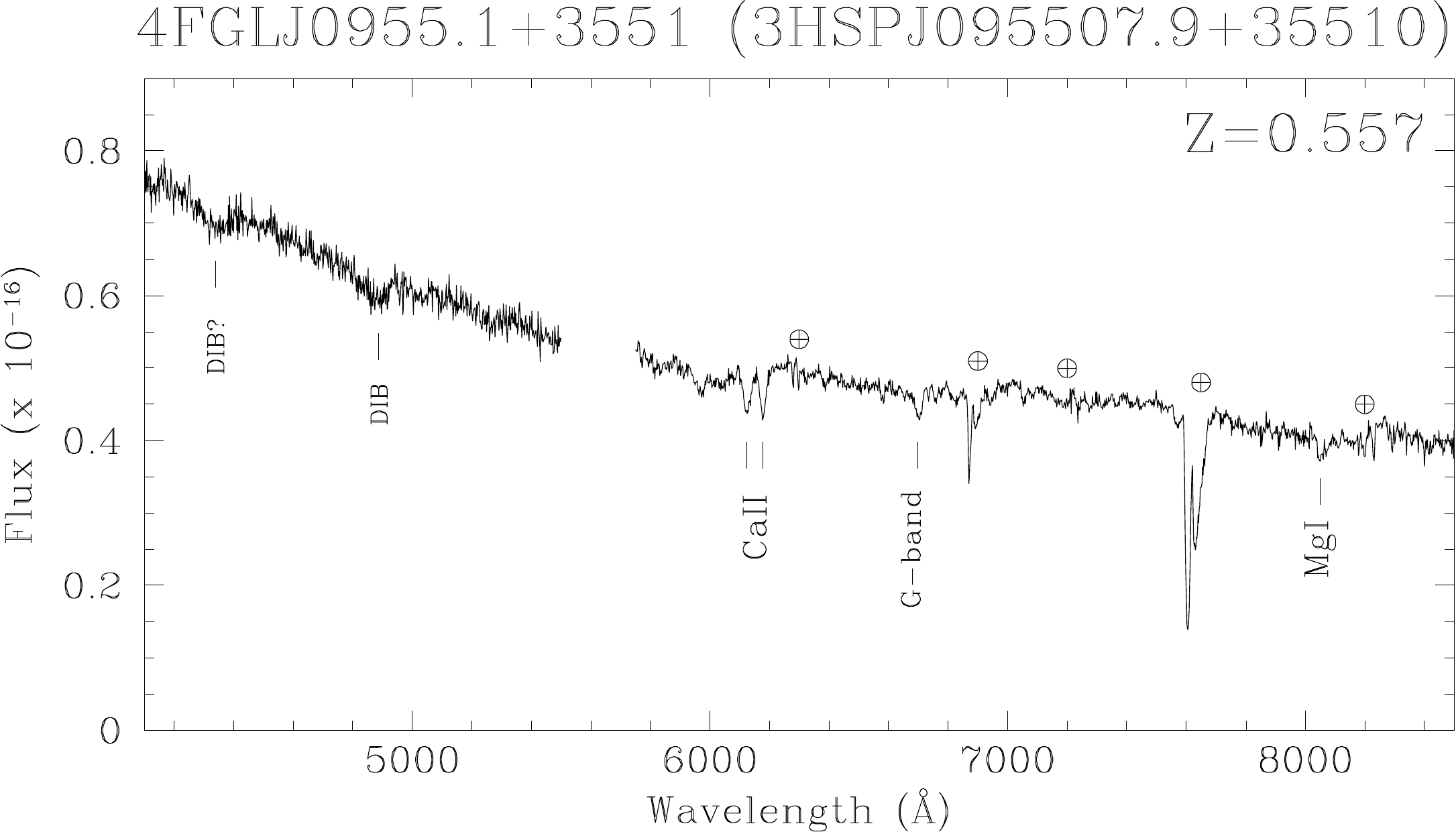}
\caption{Optical spectrum of the BL Lac object J0955 obtained at LBT+MODS (R~$\sim$~1000). The spectrum is absolute-magnitude calibrated with the magnitude of the source at the time of the observation (r~=~19.4).  The shape of the spectrum is dominated by non-thermal emission and we clearly detect three absorption lines due to the stellar population of the host galaxy at z~=~0.557.
Absorption features due to diffuse interstellar bands are labeled as DIB. The main telluric bands are marked by $\oplus$. } 
\label{fig:spectrum}
\end{figure*}

The spectrum (Fig.~\ref{fig:spectrum}) has a S/N$\sim$55 and S/N$\sim$70 for the blue and red part, respectively, and is characterized by non-thermal emission with the power-law shape (F$_{\lambda}\propto\lambda^{\alpha}$) with spectral index of $\alpha\sim$-1.1.

The spectrum clearly exhibits absorption features identified as due to Ca~II doublet, G-band~4305, Mg~I~5175 from the host galaxy (see Tab.~1 for the wavelength, ID, and the equivalent width). This allows us to derive the firm redshift of 0.557. 

We computed the nominal EW sensitivity in six intervals of the spectrum, avoiding the prominent telluric absorption features \citep[see details about the procedure in Appendix A of ][]{paiano2017tev}.  
This translates into an average minimum (3$\sigma$ level) detectable EW$\sim$0.3~$\textrm{\AA}$.
Based on the minimum EW~=~0.35~$\textrm{\AA}$, estimated in the spectral region where the emission lines of [O~II] and [O~III] are expected, at the source redshift and assuming a FWHM of 500~km/s we find that the line luminosity limits are $<$~2.2$\times$10$^{40}$~erg s$^{-1}$ and $<$~1.8$\times$10$^{40}$~erg s$^{-1}$, respectively. 


\begin{table}
\centering
\begin{tabular}{lll}
\hline 
$\lambda$  ($\textrm{\AA}$)   &  EW   ($\textrm{\AA}$)   & ID\\
\hline
6125.0  &   3.5$\pm$ 0.2  &    Ca~II~3934~$\textrm{\AA}$\\
6178.5  &   2.7$\pm$ 0.3  &    Ca~II~3968~$\textrm{\AA}$\\
6702.5  &   1.3$\pm$ 0.2  &    G-band~4305~$\textrm{\AA}$\\
8050.5  &   1.6$\pm$ 0.2  &    Mg~I~5175~$\textrm{\AA}$\\
\hline
\end{tabular}
\caption{Optical spectral features found in the J0955 spectrum: \textit{Col.1}: Central wavelength of the feature; \textit{Col.2}: Equivalent width of the feature, \textit{Col.3}: Identification of the line. }\label{tab:table1}
\end{table}

\subsection{Imaging}

\null
\begin{figure}
\center
\includegraphics[width=0.39\textwidth]{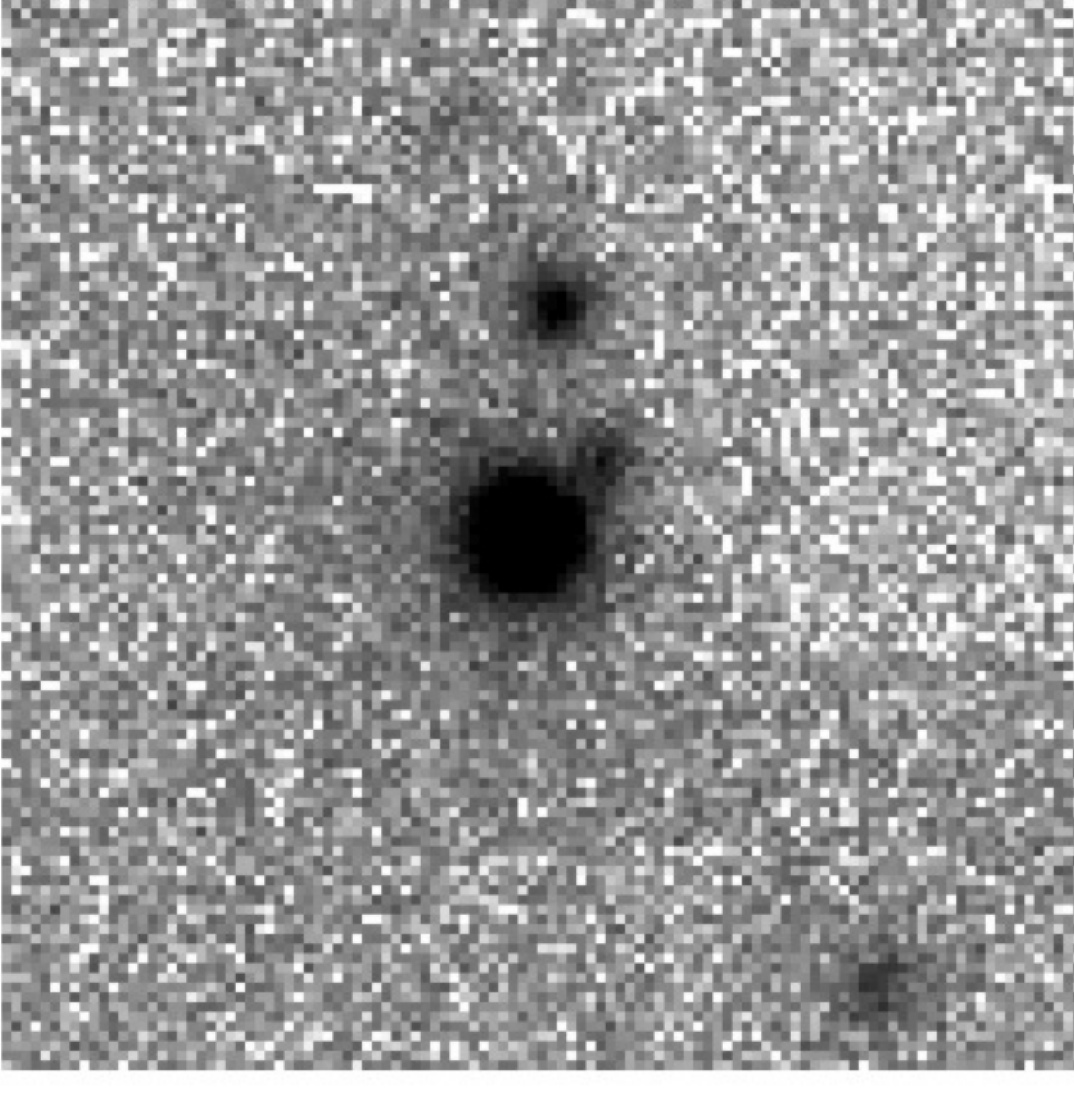}
\caption{The r band image of the BL Lac object J0955 observed by LBT and MODS1 (central brightest source). The are two resolved close companions at 1.4$^{\prime\prime}$ and 2.9$^{\prime\prime}$ from the target. Field shown is 12$^{\prime\prime}$.}
\label{fig:imaging}
\end{figure}

\begin{figure}
\center
\includegraphics[width=0.42\textwidth]{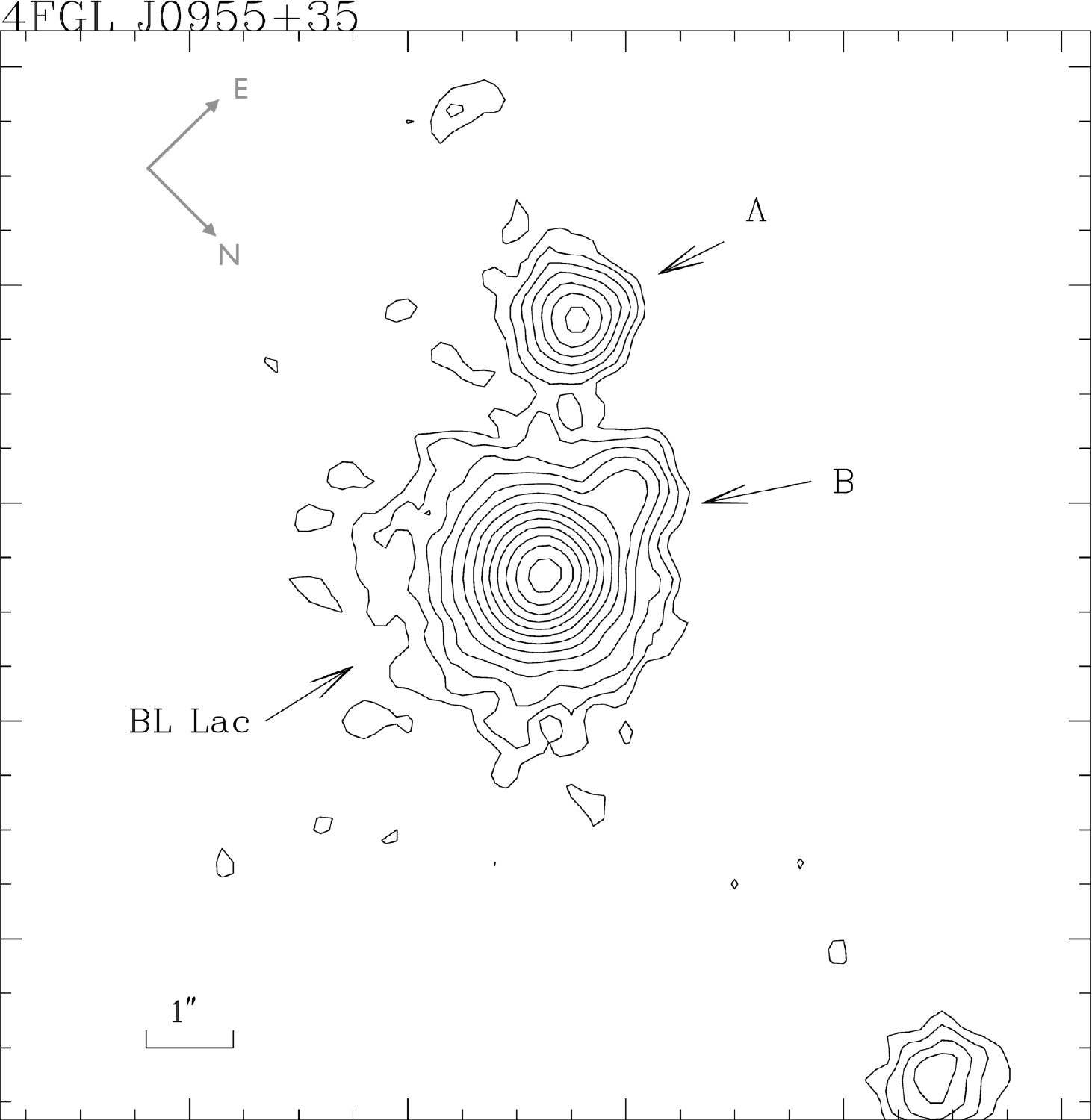}
\caption{Contour plot of the field around the BL Lac object 4FGL~J0955.1+3551 observed by LBT and MODS1 (filter r; 60 sec). There are two resolved companions: A (r~=~21.8) at 2.9$^{\prime\prime}$ (18~kpc projected distance) and B (r~=~22.5) at 1.4$^{\prime\prime}$ (9~kpc projected distance). } 
\label{fig:contour}
\end{figure}

The optical image of the target is shown in Fig.~\ref{fig:imaging}. The source magnitude at the time of the observations was r~=~19.4$\pm$0.2. 
There are two extended companion objects close to the target (see Figure \ref{fig:contour}). 
One (A) of r~=~21.8 at 2.9$^{\prime\prime}$ and a closer one (B) of r$\sim$22.5 at 1.4$^{\prime\prime}$. 
At the redshift of J0955 (z~=~0.557) these companions are at projected distance of 18~kpc and 9~kpc.  
The occurrence of close companion galaxies is not unusual for BL Lac objects \citep[e.g. ][]{falomo1996,falomo2014rev} and in a number of cases it was  found that they are physically associated with the BL Lac object \citep{pesce1994, sbarufatti2006}. In the case of J0955 no redshift information are yet available for these companions.

\begin{figure}
\center
\includegraphics[width=0.45\textwidth]{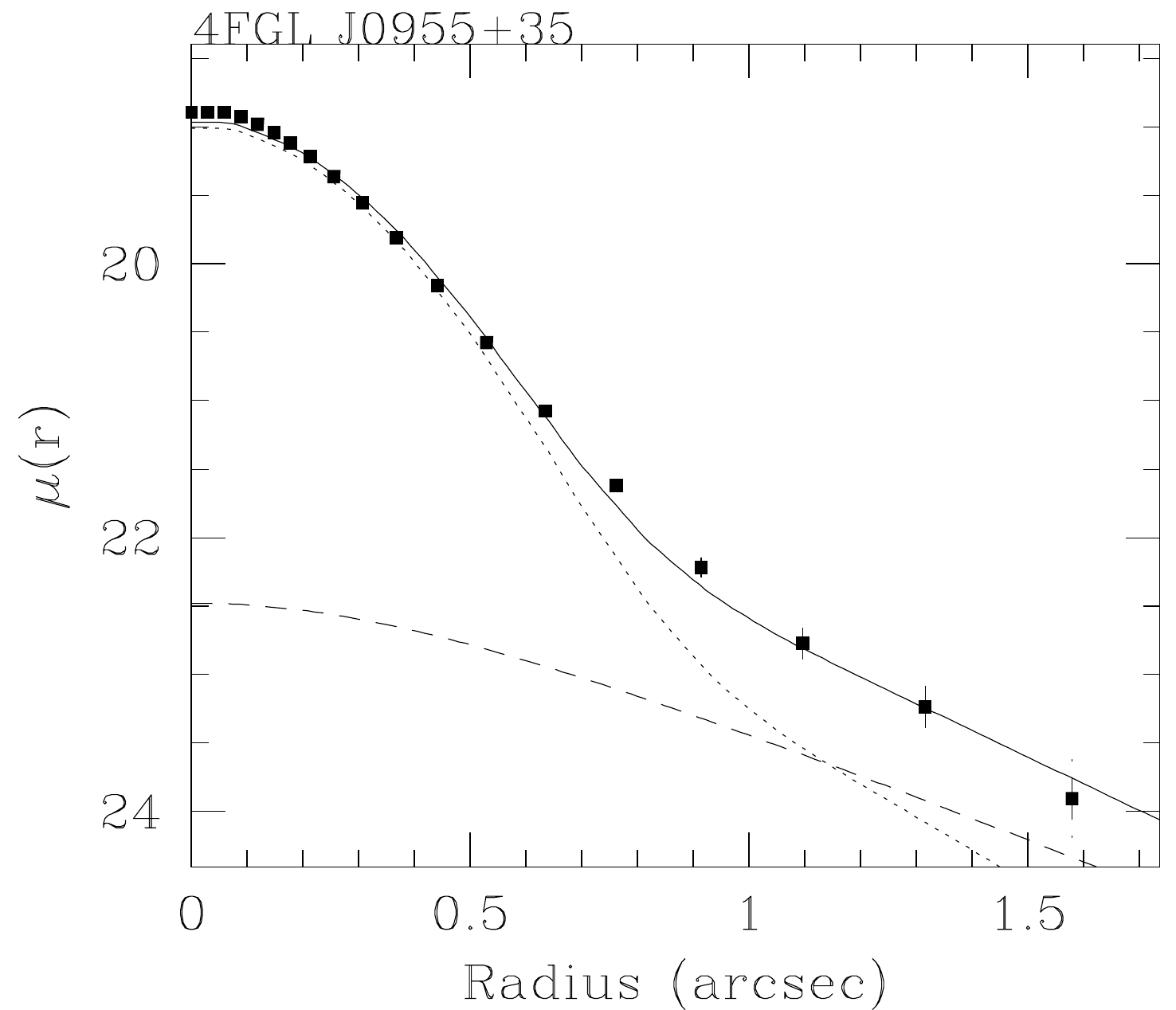}
\caption{The average radial brightness profile of the BL Lac object J0955 (filled squares) is decomposed into the nucleus (described by the PSF; dotted line) and the host galaxy (Sersic profile convolved with the proper PSF; dashed line). The best fit is given by the solid line.} 
\label{fig:fit}
\end{figure}

After masking out the companions, we performed a 2-dimensional fit of the source (see Sect. 2) and found it is well resolved (see Fig.~\ref{fig:fit}). 
The host galaxy turned out to have a magnitude r~=~20.7$\pm$0.3 and an effective radius R$_e$~=~1.2$^{\prime\prime}$. 
At the redshift of the source these correspond to an absolute magnitude M(R)~=~-22.9 and R$_{e}$~=~8~kpc,  
very typical values for the host galaxies of BLL objects \citep{sbarufatti2005}.

\section{Conclusions}

We secured high quality optical spectroscopy and imaging of the BLL J0955, proposed as the counterpart of the neutrino event by IceCube observations \citep[][]{icecube2020GCN0955}.
From the spectrum we are able to clearly determine the redshift z~=~0.557 of the source from absorption features of its host galaxy. 
At this redshift the $\gamma$-ray luminosity turns out to be $L_{\gamma} \sim 1 - 4 \times 10^{45}$ erg s$^{-1}$ (see the accompanying paper by \citealt{Giommi_2020c} for details).
No emission line are detected in the observed spectral range down to an equivalent width sensitivity of 0.3 $\textrm{\AA}$ corresponding to line luminosities $\lesssim$2$\times$10$^{40}$~erg s$^{-1}$, i.e.  
an order of magnitude fainter than those measured for TXS~0506+056 ($\sim$2$\times$10$^{41}$ erg s$^{-1}$).    

In spite of the short exposure time of the image we are also able to resolve the object and characterize the properties of its host galaxy ( M(R)~=~-22.9) that turned out to be typical for its class.
We estimate the mass of the central black hole (BH) \textit{M}$_{\rm BH}$ $\sim$3$\times$10$^{8}$ \textit{M}$_{\odot}$ using the relation between the masses of the supermassive BHs and the absolute magnitude of ellipticals \citep{bettoni2003}. This value is characteristic of the BH mass of the BLL class \citep{falomo2003}.

\section*{Acknowledgments}
\footnotesize{This work is based on observations made with the Large Binocular Telescope through a program allocated under Director's Discretionary Time.
We are grateful to the LBT-Italia director for approving the program
and to the LBTO staff (in particular Steve Allanson) for the excellent support.
The LBT is an international collaboration among institutions in Italy, the
United States, and Germany. LBT Corporation partners are: Istituto Nazionale di Astrofisica, Italy; The University of Arizona on behalf of the Arizona university system; LBT Beteiligungsgesellschaft, Germany, representing the Max-Planck Society, the Astrophysical Institute Potsdam, and Heidelberg University; The Ohio State University; and The Research Corporation on behalf of The University of Notre Dame, University of Minnesota, and University of Virginia.
A. Rossi acknowledges support from Premiale LBT 2013.
}

\label{lastpage}

\bsp	

\end{document}